# International AI Safety Report

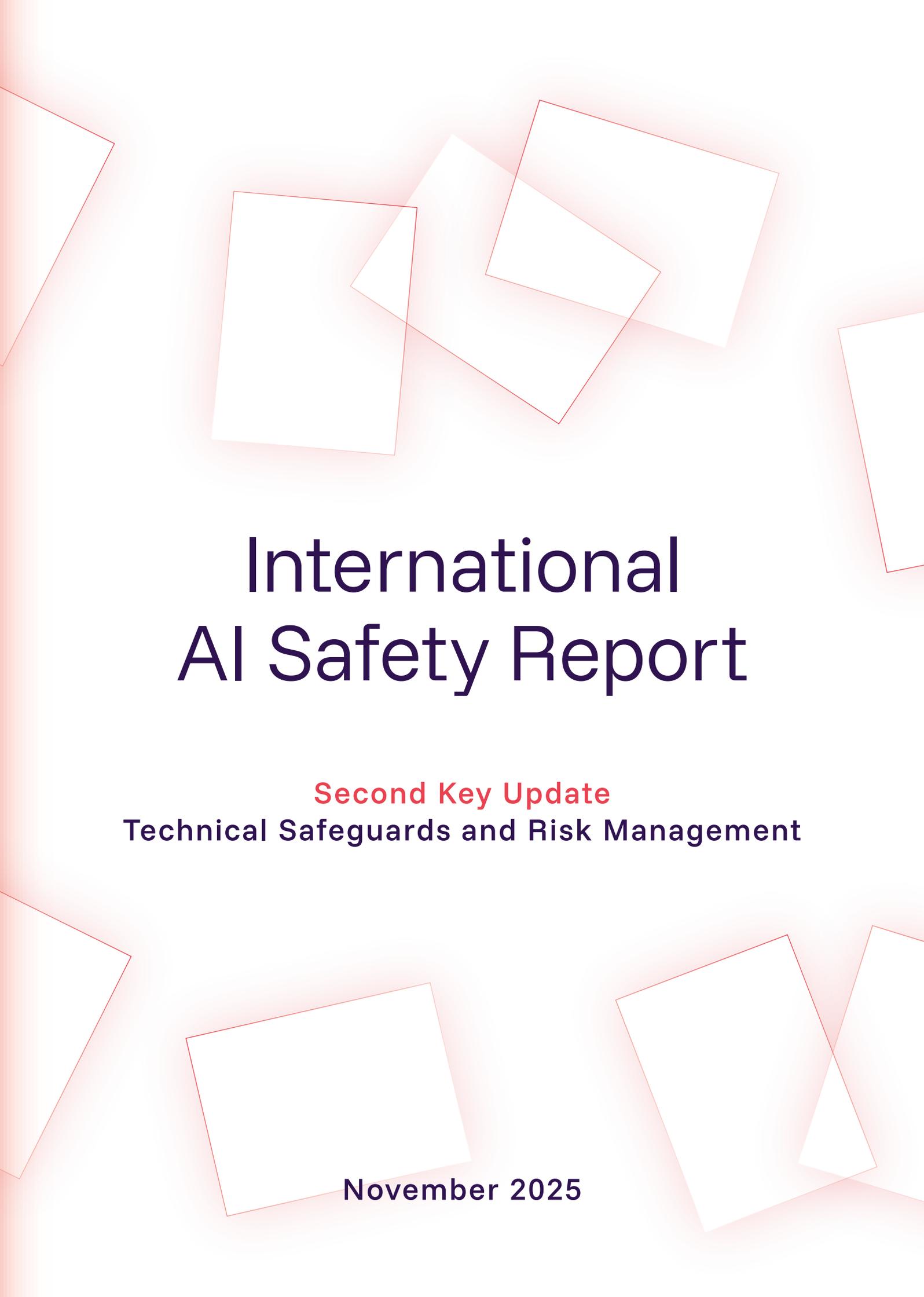

## Second Key Update
Technical Safeguards and Risk Management

November 2025

# Contributors

## Chair

**Prof. Yoshua Bengio**, Université de Montréal / LawZero / Mila - Quebec AI Institute

## Expert Advisory Panel

The Expert Advisory Panel is an international advisory body that advises the Chair on the content of the Report. The Expert Advisory Panel provided technical feedback only. The Report – and its Expert Advisory Panel – does not endorse any particular policy or regulatory approach.

The Panel comprises representatives from 30 countries, the United Nations (UN), the European Union (EU), and the Organisation for Economic Co-operation and Development (OECD). Please find here the membership of the Expert Advisory Panel to the 2026 International AI Safety Report.

## Lead Writers

**Stephen Clare**, Independent

**Carina Prunkl**, Inria

## Writing Group

**Maksym Andriushchenko**, ELLIS Institute Tübingen

**Ben Bucknall**, University of Oxford

**Philip Fox**, KIRA Center

**Nestor Maslej**, Stanford University

**Conor McGlynn**, Harvard University

**Malcolm Murray**, SaferAI

**Shalaleh Rismani**, Mila - Quebec AI Institute

**Stephen Casper**, Massachusetts Institute of Technology

**Jessica Newman**, University of California, Berkeley

**Daniel Privitera (Interim Lead 2026)**, KIRA Center

**Sören Mindermann (Interim Lead 2026)**, Independent

## Secretariat

**UK AI Security Institute:** Arianna Dini, Freya Hempleman, Samuel Kenny, Patrick King, Hannah Merchant, Jamie-Day Rawal, Jai Sood, Rose Woolhouse

**Mila - Quebec AI Institute:** Jonathan Barry, Marc-Antoine Guérard, Claire Latendresse, Cassidy MacNeil, Benjamin Prud'homme

## Senior Advisers

**Daron Acemoglu**, Massachusetts Institute of Technology

**Thomas G. Dietterich**, Oregon State University

**Fredrik Heintz**, Linköping University

**Geoffrey Hinton**, University of Toronto

**Nick Jennings**, Loughborough University

**Susan Leavy**, University College Dublin

**Teresa Ludermir**, Federal University of Pernambuco

**Vidushi Marda**, AI Collaborative

**Helen Margetts**, University of Oxford

**John McDermid**, University of York

**Jane Munga**, Carnegie Endowment for International Peace

**Arvind Narayanan**, Princeton University

**Alondra Nelson**, Institute for Advanced Study

**Clara Neppel**, IEEE

**Sarvapali D. (Gopal) Ramchurn**, Responsible AI UK

**Stuart Russell**, University of California, Berkeley

**Marietje Schaake**, Stanford University

**Bernhard Schölkopf**, ELLIS Institute Tübingen

**Alvaro Soto**, Pontificia Universidad Católica de Chile

**Lee Tiedrich**, University of Maryland / Duke

**Gaël Varoquaux**, Inria

**Andrew Yao**, Tsinghua University

**Ya-Qin Zhang**, Tsinghua University

## Acknowledgements

The Secretariat and writing team appreciated the support, comments and feedback from Markus Anderljung, Tobin South, and Leo Schwinn, as well as the assistance with quality control and formatting of citations by José Luis León Medina and copyediting by Amber Ace.

**Disclaimer**

This Update is a synthesis of the existing research on technical and non-technical risk management practices. The Update does not necessarily represent the views of the Chair, any particular individual in the writing or advisory groups, nor any of the governments that have supported its development. The Chair of the Report has ultimate responsibility for it and has overseen its development from beginning to end.

Research series number: DSIT 2025/042

# Foreword

This is the Second Key Update to the 2025 International AI Safety Report. The First Key Update (*1*) discussed developments in the capabilities of general-purpose AI models and systems and associated risks. This Key Update covers how various actors, including researchers, companies, and governments, are approaching risk management and technical mitigations for AI.

The past year has seen important developments in AI risk management, including better techniques for training safer models and monitoring their outputs. While this represents tangible progress, significant gaps remain. It is often uncertain how effective current measures are at preventing harms, and effectiveness varies across time and applications. There are many opportunities to further strengthen existing safeguard techniques and to develop new ones.

This Key Update provides a concise overview of critical developments in risk management practices and technical risk mitigation since the publication of the 2025 AI Safety Report in January. It highlights where progress is being made and where gaps remain. Above all, it aims to support policymakers, researchers, and the public in navigating a rapidly changing environment, helping them to make informed and timely decisions about the governance of general-purpose AI.

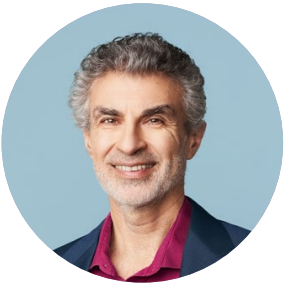

**Professor Yoshua Bengio**
Université de Montréal / LawZero /
Mila – Quebec AI Institute & Chair

# Highlights

- **Developers are adopting a strategy known as *defence-in-depth* to limit the impact of individual safeguard failures.** Combining safeguards across training, deployment, and post-deployment monitoring stages strengthens overall protection against misuse and malfunction.

- **Methods for training models to resist malicious attacks continued to improve in 2025, with prompt-injection attack success rates declining over time.** However, tests show that sophisticated attackers can still bypass safeguards around half of the time when given 10 attempts.

- **Open-weight models lag less than a year behind leading closed-weight models, shifting the risk landscape.** Greater openness can support transparency and innovation, but also makes it harder to control how models are modified and used.

- **As few as 250 malicious documents inserted into training data can allow attackers to trigger undesired model behaviours with specific prompts.** Some research shows that such data poisoning attacks require relatively few resources to carry out, regardless of model size.

- **Researchers have developed new tools to mark, identify, and track AI-generated content, though implementation remains inconsistent.** When properly used and consistently applied, techniques such as watermarking, content detection, and model identification can help researchers study the spread of AI models and systems, trace their outputs, and verify human-created content.

- **The number of AI companies with Frontier AI Safety Frameworks more than doubled in 2025: at least 12 companies now have such frameworks.** These frameworks describe tests and risk mitigation measures that AI developers intend to implement as their models become more capable.

# Introduction

This second Update to the 2025 International AI Safety Report assesses new developments in general-purpose AI risk management over the past year. It follows the First Key Update, which examined new evidence on AI capabilities and risks. The First Key Update highlighted AI progress on mathematical reasoning, coding, and autonomous operation, and discussed emerging biological risks, cyber offence risks, and challenges in overseeing increasingly advanced models (*1*). This Update examines how researchers, public institutions, and AI developers are approaching risk management for general-purpose AI, including policies and techniques to make general-purpose AI more reliable and resistant to misuse.

This work has taken on new significance as recent capability advances have prompted more robust risk management. For example, in recent months, three leading AI developers† applied enhanced safeguards to their new models, as their internal pre-deployment testing could not rule out the possibility that these models could be misused to help create biological weapons (*1*). As a precautionary measure, these developers reported adopting stronger protections, including enhanced security measures, deployment controls, and real-time monitoring (*2\**, *3\**, *4\**).

Beyond these specific precautionary measures, during the past year there have been a range of other advances in techniques for making AI models and systems more reliable and resistant to misuse. These include new approaches in adversarial training, data curation, and monitoring systems. In parallel, institutional frameworks that operationalise and formalise these technical capabilities are starting to emerge: the number of companies publishing Frontier AI Safety Frameworks‡ more than doubled in 2025, and governments and international organisations have established a small number of governance frameworks for general-purpose AI, focusing largely on transparency and risk assessment.

This Update examines these developments in two parts. Section 2 reviews technical methods for improving robustness and reliability throughout the AI lifecycle, from model training to post-deployment monitoring. Section 3 examines how institutional frameworks are beginning to turn these technical approaches into formal guidelines, including through transparency requirements, evaluation standards, and risk management procedures. Throughout, the Update documents what has changed since the 2025 International AI Safety Report was published and identifies key uncertainties about current approaches.

---

†   Anthropic, OpenAI, and Google DeepMind.

‡   Sometimes also called 'Frontier AI Frameworks' or 'Safety and Security Frameworks'.

# Technical methods for improving reliability and preventing misuse

## Key information

— **Researchers have refined training methods that make models more reliable and resistant to misuse.** Improved techniques correct biased human feedback and provide evaluators with tools to detect errors. Their effectiveness varies across deployment settings and use-cases. The broader attack-defence landscape remains dynamic, as sophisticated adversaries continue to find ways to bypass defences.

— **Developers and deployers can identify and prevent some undesired behaviours by monitoring the behaviour of AI models and systems, but these methods have substantial limitations.** Monitoring can be applied to hardware, input prompts, internal computations, and model outputs.

— **New provenance and watermarking tools can help trace AI-generated content.** Yet these signals can still be removed, forged, or degraded through relatively simple post-processing or manipulation.

— **Since individual techniques have limitations, AI developers implement multiple safeguards, a strategy known as *defence-in-depth*.** Layering safeguards means that harmful events can be prevented even if one layer fails.

Technical safeguards constitute one strand of general-purpose AI risk management. These mechanisms typically support two related goals: preventing misuse (for example by training models to refuse dangerous requests), and preventing malfunctions (for example by detecting when models produce factually incorrect outputs) or limiting the potential damage they cause.

2025 has seen continued research into techniques to train models to refuse harmful requests, prevent dangerous capabilities from emerging, and maintain human control over increasingly autonomous systems. However, important evidence gaps regarding their real-world effectiveness remain (*5*, *6*). These gaps result in part from the rapid pace of model development and deployment, which makes it difficult to evaluate safeguards under realistic conditions and collect systematic data on their effectiveness. Safeguards are developed amid continual changes in how AI models and systems are built, what is known about how they work, and how attackers or malicious users seek to misuse them. This shifting threat landscape suggests a need for continual developing, testing, and refining of safeguards.

This section examines developments for technical safeguards across three stages of the AI development lifecycle:

— **Model training:** Methods that are applied during training and design, such as giving models specific kinds of feedback (*7*) or stopping them from generating harmful responses (*8*).

— **Product deployment:** Methods that are applied when one or more AI models are integrated into products, such as tools that detect harmful outputs (*9\**) or attempts to circumvent safeguards (*10*).

— **Post-deployment monitoring:** Tools used to monitor how AI systems are being used after deployment, such as watermarking to trace AI-generated content.

These methods can be made more robust by implementing multiple safeguards in layers, a principle known as *defence-in-depth* (Figure 1). Using multiple safeguards in sequence helps reduce the chance of harm: if one safeguard fails the others may still succeed. Figure 1 shows how this approach can span the AI lifecycle:

— The first layer of defence might involve **training interventions**, such as reinforcement learning from human feedback (discussed below) or other safe-by-design methods, to limit undesired behaviours early in development.

— The second layer adds **deployment interventions**, such as classifiers or guardrails that make it harder for users to generate potentially harmful outputs.

— After deployment, a third layer of **post-deployment monitoring** tools, such as watermarking and content-provenance systems, can help detect misuse.

— The final layer (mentioned here for comprehensiveness) extends beyond technical measures to include societal **resilience measures**: the ability of societal systems to resist, recover from, or adapt to harms. This Update does not discuss resilience measures, though they will be addressed in the forthcoming 2026 International AI Safety Report.

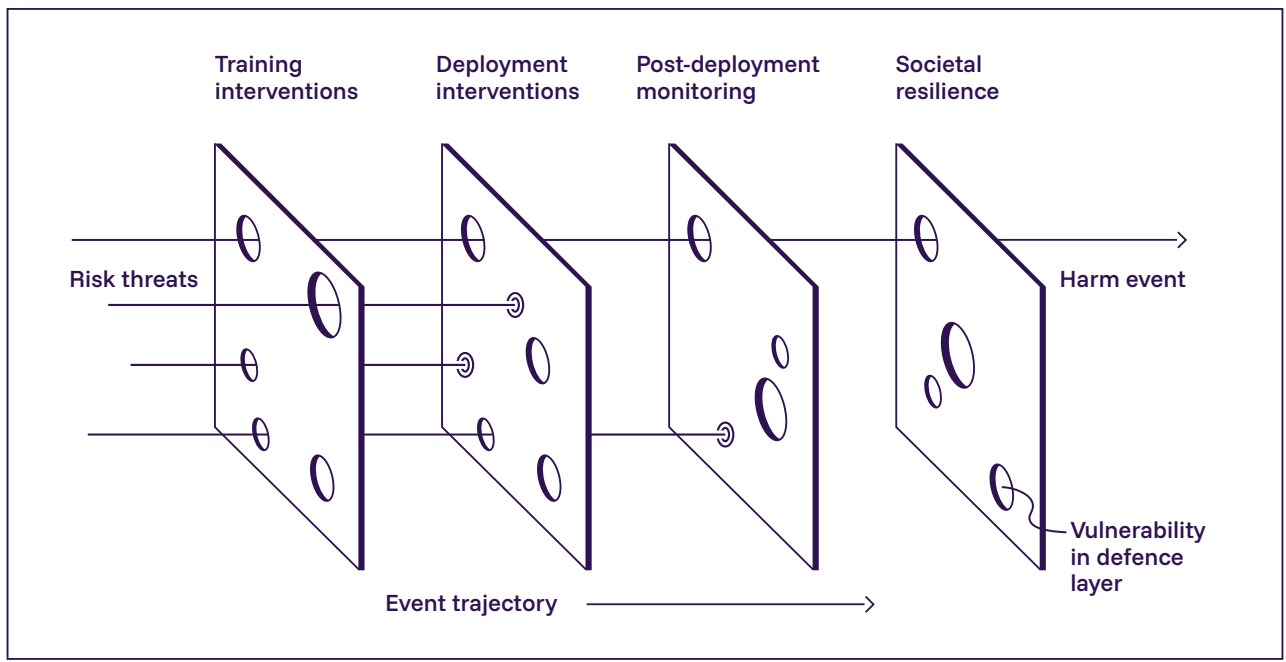

**Figure 1:** A 'Swiss cheese diagram' illustrating the defence-in-depth approach: multiple layers of defences can compensate for flaws in individual layers. Current risk management techniques for AI are all flawed, but layering them can offer much stronger protection against risks.

# Training safeguards: limiting undesired behaviours

## Training techniques can sometimes prevent models from developing harmful capabilities

One approach to making AI models and systems more resistant to misuse is to prevent them from developing undesired capabilities in the first place by removing harmful knowledge from the pre-training data (*11*, *12\**, *13*). These techniques show promise, though emerging evidence suggests that they may be more effective at preventing complex harmful behaviours, such as assisting in weapon development (*13*), than at eliminating simpler undesired capabilities, like generating offensive text (*14*). It is also difficult to ensure that all instances of harmful training material have been removed from training data, given the large size of the datasets developers use to train leading models (*15*). While broad links between dataset size/diversity and overall model performance are well established, there remains uncertainty on how specific data characteristics and training dynamics influence the emergence of new capabilities and behaviours in large models (*16*). Moreover, there is a trade-off between safety and usefulness, and efforts to limit the development of certain capabilities can hinder commercial objectives (*17*, *18*). For example, AI systems with strong coding capabilities are highly valuable for legitimate use cases but could also be misused for offensive cyber operations.

## Ensuring AI models are reliable and resistant to misuse remains difficult

While preventing harmful capabilities from arising at all is one approach, developers also train models to resist misuse even when they possess potentially dangerous knowledge. Relevant approaches here include training models to refuse harmful requests, provide truthful information, and decline tasks beyond their capabilities. Research has continued to advance these goals. However, it remains difficult to specify desired behaviours precisely enough for models to reliably exhibit them across the wide range of real-world uses. When they fail, models may produce harmful content, follow dangerous instructions, or behave unpredictably.

One common approach for training AI models to exhibit desirable behaviour is called *reinforcement learning from human feedback* (RLHF) (*19\**). It involves human evaluators rating model outputs and training the model to learn from these ratings. The technique is well-established, but it suffers from the fact that human feedback can be inconsistent, systematically flawed, or incomplete (*20*, *21*). In 2025, RLHF methods have continued to evolve, as researchers have refined how human feedback is collected, interpreted, and applied. Recent work has explored ways to detect and correct misleading patterns in human feedback that might reduce training effectiveness (*22*). Other research has focused on improving the quality of the feedback itself by giving evaluators – or AI models themselves – tools to better detect and correct errors in a model's responses (*23\**, *24*). In parallel, open source initiatives are releasing datasets, code, and training recipes for new improvements in RLHF, expanding transparency, reproducibility, and shared experimentation across the research community (*25\**, *26*).

## Sophisticated attackers routinely find ways to elicit harmful behaviours from AI models

As the capabilities of general-purpose AI models advance, the potential for misuse also grows. An important approach for mitigating misuse risk is *adversarial training* (*27*), in which developers create 'attacks' that attempt to elicit harmful behaviours and then train the model to resist them by refusing inappropriate requests.

In 2025, researchers have improved adversarial training, including by developing new methods to scale the technique more effectively (*28*, *29\**, *30\**, *31*, *32*) and algorithms that make models more robust to attacks (*8*, *33*, *34*, *35*, *36*). Yet adversarial training remains imperfect: attackers routinely devise new attacks that succeed in eliciting harmful behaviour from new models (*37*, *38\**, *39\**, *40*). For example, one recent study crowd-sourced 'prompt injection' attacks, which involve giving an AI model specific inputs designed to circumvent safeguards. The study found over 60,000 successful attacks (*39\**). The success rate of prompt injection attacks, as reported by AI developers, has been falling slightly over time, but when given ten tries, attackers can still successfully execute such attacks about half the time (Figure 2).

Some evidence suggests that the cost of circumventing safeguards is decreasing relative to the cost of developing and maintaining them. For example, recent research shows that as few as 250 malicious documents inserted into a model's training data can cause it to produce undesired outputs when given specific prompts (*42*). This suggests that launching such 'data-poisoning' attacks could require far fewer resources than building or maintaining robust defences. Other studies have investigated model fine-tuning, which involves training models with additional data to adapt them to specific tasks. These studies find that fine-tuning models for one harmful purpose can cause them to behave harmfully in other, unrelated contexts (*43\**). For example, models fine-tuned to write insecure code subsequently gave malicious advice when responding to prompts about entirely different topics (*44*). An emerging challenge is that as AI models and systems become more generally capable and able to act more autonomously (*1*), they can be deployed in more diverse environments. This creates new opportunities for attackers if safeguards developed in training contexts do not generalise to these varied real-world settings (*45, 46, 47, 48, 49, 50, 51*).

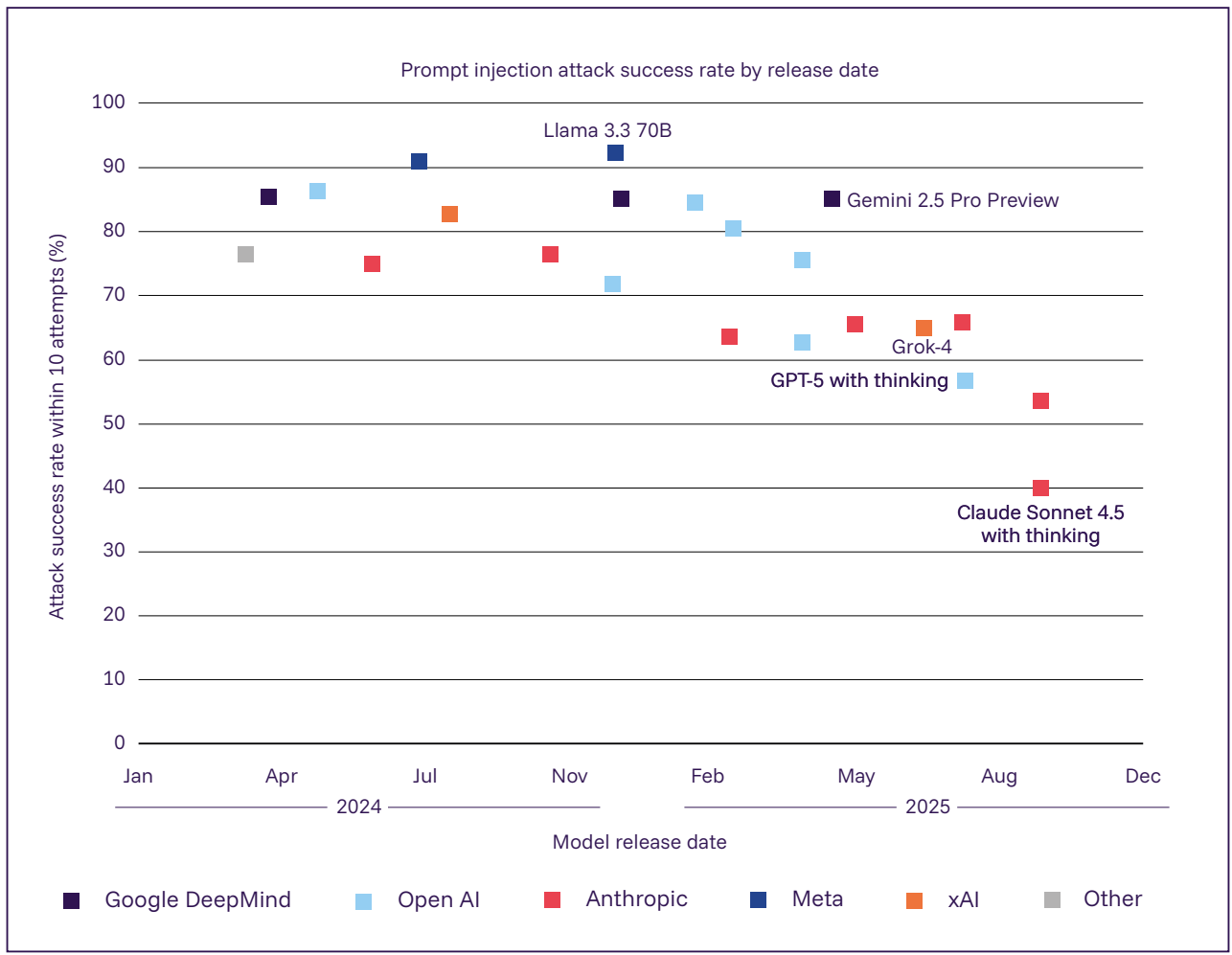

**Figure 2:** Prompt injection attack success rates, as reported by AI developers for major models released between April 2024 and July 2025. Each point represents the proportion of successful attacks within ten attempts against a given model shortly after release. The reported success rate of such attacks has been falling over time, but remains relatively high. Source: Zou et al. 2025 (*39\**), cited in Anthropic 2025 (*41\**).

**Open-weight models are rapidly advancing, but risk mitigation techniques remain immature**

Significant developments have occurred in the open-weight model ecosystem since the publication of the last Report. 'Open-weight' means that a model's weights – the parameters that determine how the model generates outputs – are freely available for download. Open-weight models' capabilities now lag behind those of leading closed-weight models by less than one year, a gap that has shortened over time (*52*, *53*). Across language (*54\**, *55\**, *56\**, *57\**, *58\**, *59*), image generation (*60\**, *61\**), and video generation (*62\**, *63\**), open-weight models offer significant benefits for research in capabilities and security, access, transparency, and reproducibility (*64*, *65*, *66\**, *67*, *68*, *69*, *70*). Importantly, open- and closed-weight ecosystems influence each other. Techniques and risk mitigation practices developed in open-weight models are frequently adopted by proprietary developers, while improvements in closed systems often inform open source research.

However, the availability of open-weight models also affects the risk landscape (*65*, *67*, *70*, *71*, *72*). Because they can be freely downloaded, open-weight models can be used and modified without oversight and control by the initial developer (*73*). For example, modified open-weight image-generation models have become the most common tools used for creating AI-generated child sexual abuse material (*74*, *75*, *76*). Analyses of open-weight ecosystems show that numerous open-weight models have been fine-tuned to specifically perform harmful tasks or disable safeguards, highlighting ongoing challenges in monitoring and mitigating downstream misuse (*77*).

Research is ongoing into techniques to make open-weight models more resistant to misuse and fine-tuning for harmful purposes (*66\**, *69*, *78*, *79*). For example, researchers and providers have been exploring 'unlearning' techniques, which aim to make models resistant to harmful modification (*8*, *35*, *80*, *81\**, *82*, *83*, *84*). However, recent research shows that these techniques can be reversed by fine-tuning the model on fewer than 100 examples (*73*, *85\**, *86*, *87*, *88\**, *89*). Complementary research explores ways of directly modifying how models process and represent concepts to suppress harmful knowledge (*90*, *91*). This could include, for example, removing concepts such as 'violence' or 'abuse' from a model's internal representation (*90*). While this can help resist some forms of misuse, these modifications can often be reversed by actors with the technical skill to fine-tune models. Other recent evidence suggests that filtering harmful topics from pre-training data can more robustly prevent misuse of open-weight models (*13*). Overall, though, technical risk mitigation techniques for open-weight models remain immature (*77*).

# Deployment safeguards: monitoring and preventing potentially harmful behaviours

While developers apply training-based methods during model development, they apply deployment-based safeguards when the model is evaluated, integrated, or used within broader systems or products. These include classifiers, filters, or monitors.

**Monitoring and intervention tools can detect and prevent many potentially harmful behaviours**

A key intervention is to monitor AI systems for signs of risky behaviour and intervene before they cause harm. Actors developing or deploying AI systems can implement system monitoring at multiple points (Figure 3), including:

— **Hardware and system-level compute:** Monitor computational resources used to train or run AI models and systems, including hardware configurations and operational environments, to verify that they are developed and deployed under appropriate conditions (*92*, *93*).

— **Inputs:** detect or flag suspicious or potentially-harmful requests (*9\**, *94*, *95*, *96\**).

— **Internal computations:** observe the AI model's *internal* activities to identify early signs of unsafe behaviour (*97*, *98*).

- **Chain of thought monitors:** review the intermediate reasoning steps that some AI systems generate before producing a final answer (*99\**, *100\**).
- **Outputs:** identify potentially harmful AI-generated content (*96\**, *101*).

Monitoring and intervention techniques can prevent many potential harms (*9\**, *102*, *103*). However, recent research has also highlighted new challenges and limitations. For example, monitoring a model's chain of thought can help developers and researchers understand why it generated a potentially harmful response; but giving the model feedback based on information in its chain of thought, may lead it to conceal suspicious reasoning steps while still generating undesired outputs (*98*, *100\**, *104*). Other research has shown that even multiple layers of safeguards can be vulnerable to sophisticated attacks that are specifically designed to break each layer (*96\**).

Another emerging challenge is that of overseeing more autonomous AI systems that can initiate or execute actions on behalf of users. As such systems become more capable and operate in more diverse environments, oversight can become more challenging due to the speed with which they can act and the complexity of the environment. Some developers are now implementing human-in-the-loop controls that require users to explicitly confirm an agent's plans before it takes action (*105\**, *106*, *107\**).

Developers can also act after identifying potentially risky behaviours. For example, they can log information, filter or modify harmful content, flag abnormal activity, and trigger failsafes or human overrides (*108*). Some AI developers have adopted a range of monitoring processes to detect and respond to such events (*109*, *110\**, *111\**). Monitoring the use of AI systems can also help developers meet transparency goals. Logging and incident reporting creates data about the frequency and characteristics of AI incidents, which can inform other risk management processes (*112*).

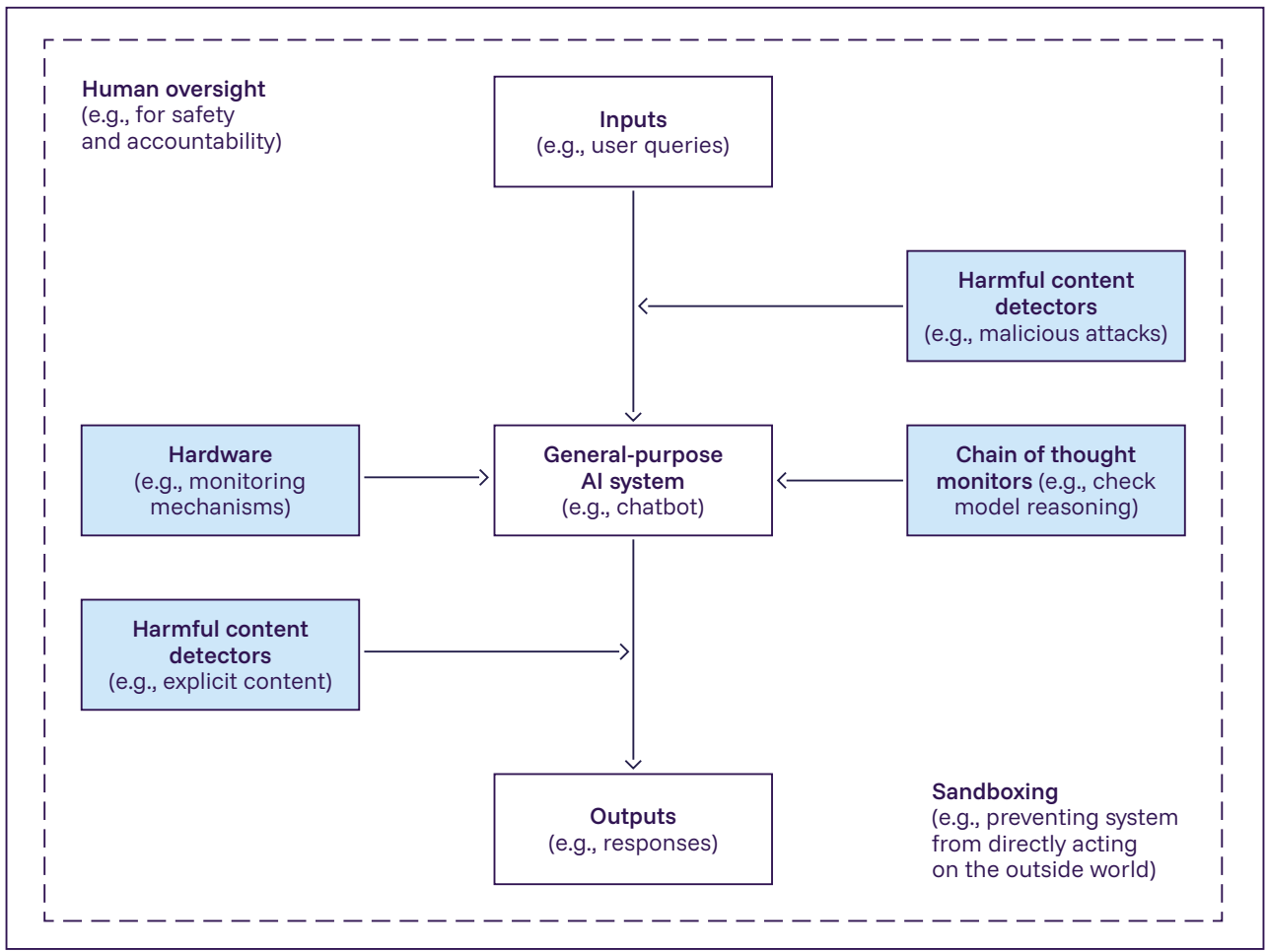

**Figure 3:** Monitoring, intervention, and control techniques can be applied to general-purpose AI system inputs, outputs, and models themselves to help researchers and deployers oversee system behaviour and establish guardrails. Source: Bengio et al., 2025 (*70*).

# Post-deployment monitoring tools: model, system, data, and output provenance

Beyond monitoring individual systems during deployment, post-deployment monitoring tools trace how AI models, agents, and their outputs circulate and are used in the real world. This helps people track the origin of AI models, systems, and AI-generated content (provenance); investigate incidents when harms occur; and implement accountability mechanisms.

## New techniques to track the use and origin of AI models and systems

AI model identification techniques help trace where and how specific AI models and systems are used. When harms occur, knowing which model was involved can inform how actors should respond. This can be particularly important for open-weight models, as they are distributed in a variety of ways – from mainstream hosting platforms such as Hugging Face to less regulated channels such as Dark-Web forums.

One tracking approach is to give AI models or systems unique identifying characteristics (*113*, *114*, *115*, *116*, *117\**). As a simple example, models can be trained to always respond with their name and version when asked, "Who are you?". But other, more subtle identifying techniques can also be applied. For open-weight models, for example, developers can embed unique watermark patterns in the model weights themselves (*114*, *118*, *119*, *120*, *121\**, *122*). Researchers are also developing new methods to infer the provenance of open-weight models that lack watermarks – for example, determining whether one model was created by fine-tuning another (*123*, *124*, *125*).

While model-tracking techniques help establish accountability and monitor misuse, they also raise potential privacy concerns. Overly broad monitoring could enable surveillance of legitimate research or user activity, creating tensions between accountability goals and individual or institutional autonomy (*126*, *127*).

## Improvements in AI content detection techniques

Watermarks, metadata, and other AI content detectors help researchers track the spread of AI-generated content, study its impacts, and identify sources when harms occur.

Work on these techniques has advanced in three main areas since the publication of the last Report. First, there have been improvements in both cryptographic provenance methods, which use digital signatures, hashes, or encryption to verify authorship, and digital 'watermarking' methods that add subtle, distinctive patterns to AI-generated content with information about its origin (see Figure 4) (*128*, *129\**, *130*). These watermarks include distinctive word choices for text outputs (*131*, *132*), subtle patterns in pixels for AI-generated images and videos (*133*), and patterns embedded in audio waves for audio outputs (*134*). Recent studies have strengthened the reliability of these approaches. For example, one study combines cryptographic fingerprinting with semantic-aware encoding to make image provenance verifiable even after compression or some forms of editing (*135*). In another study, researchers improved text watermarking by embedding statistical signals across multiple vocabulary channels while preserving fluency (*136*).

Second, standardised file formats for AI-generated content that include information about how it was generated have improved (*137*).

Finally, researchers have continued to develop AI content detectors—systems trained specifically to distinguish AI-generated from human-created content based on statistical patterns (*138*, *139*, *140*, *141\**).

Each of these techniques has useful applications but remains vulnerable to deliberate tampering. Recent studies demonstrate that watermarking and provenance signals can be removed, forged, or degraded through relatively simple post-processing or adversarial manipulation, indicating that defensive methods currently lag behind adaptive attacks (*143\**, *144*, *145*).

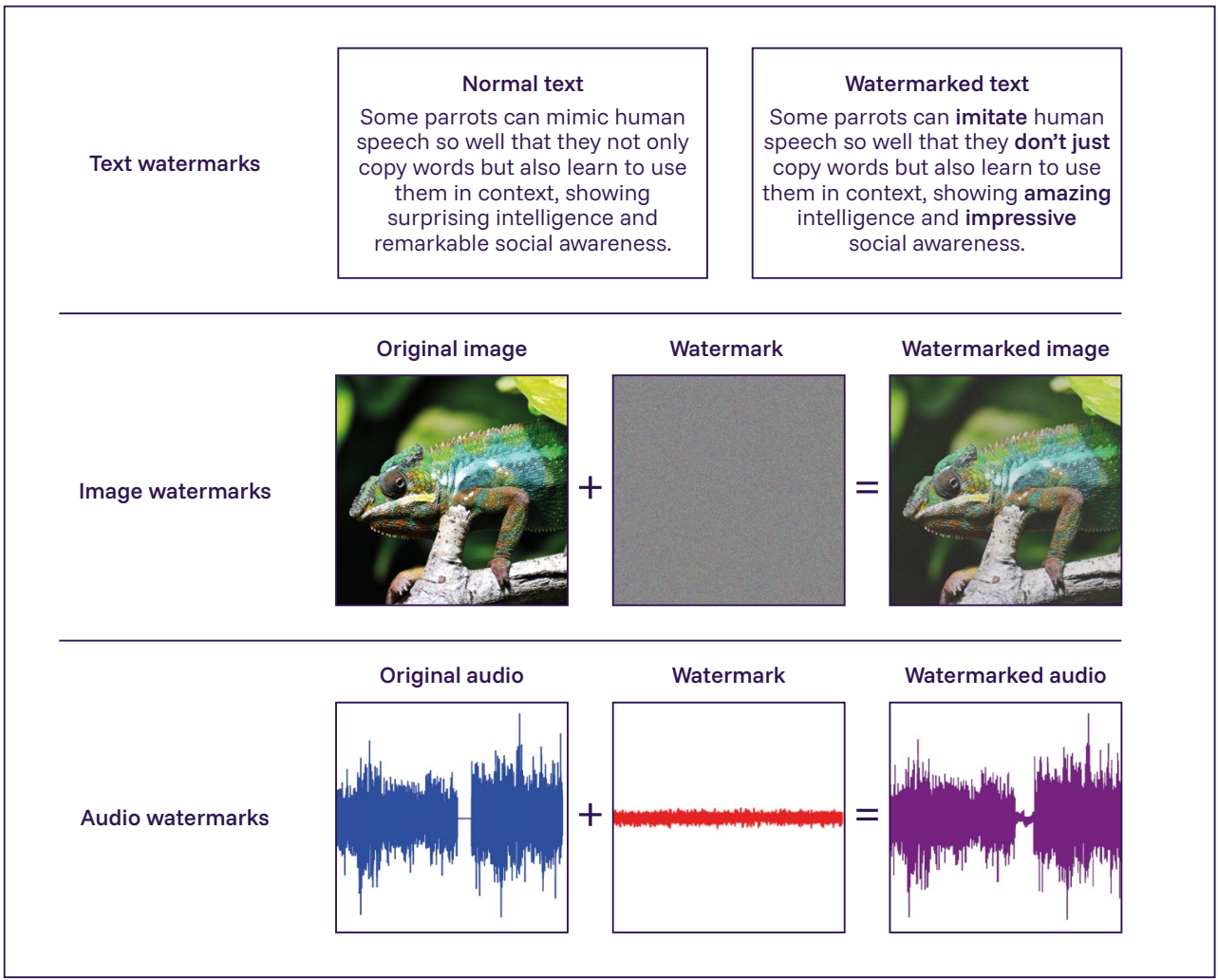

**Figure 4:** Watermarking is one of several key techniques for studying the downstream uses and impacts of AI-generated content. Source: William Warby 2024 (*142*).

## Identity and accountability mechanisms are emerging for AI agents

As AI models and systems increasingly act autonomously online – they can now make purchases, send emails, and execute code – establishing clear accountability is becoming more important. If an AI agent causes harm or violates policies, investigators need to determine which system was responsible and under whose authority it was acting.

Recent work has begun developing technical frameworks to address these challenges.

Proposed approaches include giving AI agents secure digital identities that can be verified and audited (*146*), creating unique identifiers for specific AI system instances (for example, distinguishing between different conversations with the same model) (*147*), and developing systems that allow users to set explicit permissions for what AI agents can do on their behalf while maintaining auditable logs of agent activities (*148*). However, these frameworks remain in early stages of development, and it is too early to assess whether they will prove effective or practical at scale. Real-world deployment will need to address challenges such as implementation costs, integration with existing systems, and resistance to tampering or circumvention.

# Institutional approaches to risk management

## Key information

— **Since January 2025, some governments and international organisations have begun establishing frameworks for general-purpose AI that emphasise transparency and oversight.** Examples include the EU General-purpose AI Code of Practice, China's AI Safety Governance Framework 2.0, and the G7/OECD Hiroshima AI Process Reporting Framework. These approaches have a prominent focus on transparency requirements, model evaluations, and incident reporting.

— **12 AI companies now reportedly have Frontier AI Safety Frameworks – this number has more than doubled since early 2025.** These frameworks describe the measures companies will take as they develop more capable AI models, including potentially heightened protections such as access restrictions or deployment delays.

— **Evidence-based assurance methods are beginning to complement traditional risk management.** Developers are experimenting with safety cases, incident analyses, and performance logs to document and justify AI risk identification and mitigation practices, marking a shift toward more structured and transparent assurance.

Several of the mechanisms discussed above are now being integrated into institutional frameworks. Governments, international organisations, and developers are beginning to embed technical risk identification and mitigation measures, such as model evaluations, provenance tracking, and red-teaming, within broader oversight structures. These efforts notably include transparency and reporting requirements, but also formal risk-management procedures, and internal assurance mechanisms to ensure consistent application and documentation.

This section's discussion of risk management approaches is descriptive: it aims to inform actors in the AI ecosystem about current global approaches to AI risk management. Where available, evidence on the effectiveness and limitations of these approaches is discussed, but policy recommendations are outside the scope of this work.

## Early governance approaches supplement technical risk mitigation practices

As technical risk management techniques evolve, governments, companies, and international organisations have begun translating them into governance requirements and oversight practices that specifically address general-purpose AI. Since the publication of the last Report (January 2025), several institutions have introduced or updated frameworks that formalise expectations around topics such as transparency, model evaluations, and incident reporting. Examples include the EU General-purpose AI Code of Practice (*149*), China's AI Safety Governance Framework 2.0 (*150*), and the G7/OECD Hiroshima

Institutional approaches to risk managementAI Process Reporting Framework (*151*, *152*, *153*, *154*). Similar initiatives are emerging beyond Europe and North America. For example, The ASEAN Expanded Guide on AI Governance and Ethics (Generative AI) (*155*) and South Korea's Basic Act on AI (*156*, *157*) both introduce disclosure, labelling and pre-deployment evaluation requirements for AI models.

These frameworks represent early efforts to formalise technical safeguards – such as model evaluations, red-teaming, and content provenance – into oversight requirements. However, most were introduced in 2025, making it too early to assess whether they will be effectively implemented or achieve their intended safety outcomes.

## At least 12 AI companies have adopted Frontier AI Safety Frameworks

Alongside emerging auditing and evaluation practices, the development of safety frameworks for the most advanced AI systems has accelerated significantly. At least 12 AI companies† have now published or announced 'Frontier AI Safety Frameworks' – policies describing how they plan to test, monitor, and control their most capable general-purpose models before deployment. Over half of these were announced in 2025 (*158*).

Frontier AI Safety Frameworks aim to function as risk management tools for specified risks (*159*). Currently they focus mainly on chemical, biological, radiological, and nuclear risks, cyber capabilities, and autonomous behaviour from AI models (*159*), though some developers consider additional risks as well (*160*). While they vary in scope and detail (*160*, *161*), frameworks commonly include elements such as:

— **Model evaluations:** Comprehensive evaluations designed to elicit full model capabilities and identify potential harms, ensuring risks are not underestimated.

— **Capability thresholds:** Predefined levels of dangerous capabilities that, if crossed, trigger heightened protections or restrictions.

— **Security measures:** Safeguards to prevent unauthorised access to model weights and other sensitive assets that could enable misuse or replication of dangerous capabilities.

— **Mitigations:** Options for responses to risk assessments, ranging from access restrictions to development or deployment pauses (*161*).

Many of the measures specified in Frontier AI Safety Frameworks are voluntary. This may increase adaptability, but also be a limitation. For example, one study found that AI companies have inconsistently fulfilled previous voluntary commitments, demonstrating the highest commitment to content provenance measures such as watermarking and the lowest commitment to model weight security efforts (*162*).

Another current limitation is the lack of clear methods for defining acceptable risk levels, linking capabilities to those levels, and determining appropriate mitigations. In many cases, frameworks do not explicitly link capabilities to their potential real-world impact. As a result, risk thresholds are often implicit: they are reflected in their chosen responses, such as restricting access, enhancing monitoring, or delaying deployment, rather than based on measurable evidence of harm (*163*, *164*). Frameworks vary, reflecting the early stage of the field. This diversity poses coordination and standardisation challenges, but may also reveal which approaches prove most effective over time.

## Evidence-based assurance methods are beginning to complement traditional risk management

Developers are beginning to adopt assurance tools to substantiate claims about the reliability, robustness, and governance of their general-purpose AI products. One emerging approach is the use of safety cases, adapted from other safety-critical fields such as aviation and nuclear engineering (*103*, *165*, *166*, *167*, *168*, *169*, *170*, *171*). These are structured arguments that make claims about acceptable model or

---

† The following companies have published such frameworks: Anthropic; OpenAI; Google DeepMind; Magic; Naver; Meta; G42; Cohere; Microsoft; Amazon; xAI; Nvidia.

International AI Safety Report: Second Key Update                                                                                              16 system behaviour and provide evidence that the model or system will behave within these bounds. Though their current adoption is limited and their suitability for general-purpose AI systems remains contested, several leading AI developers now reference safety cases in their risk management frameworks. For example, developers use safety cases or equivalent analysis to justify mitigations to implement when models reach specific capability thresholds (*41\**, *172\**, *173\**).

Industry adoption of safety cases is inconsistent. An OECD analysis of 20 organisations participating in the Hiroshima AI Process Reporting Framework found that most combine quantitative metrics with expert qualitative judgment across the AI lifecycle, but few disclose full methodologies or systematically validate whether documented controls correspond to real-world outcomes (*174*). Other studies have similarly found that even leading firms with formal safety frameworks often lack systems for connecting actual incidents to their safety documentation (*163*), and that evaluation reports omit details, making them hard to interpret (*175*).

## Conclusion

The technical safeguards that developers can implement throughout the AI lifecycle can reduce specific vulnerabilities, but they have significant limitations. Current risk mitigation methods can be circumvented by sophisticated actors, vary in effectiveness across different deployment contexts, and are often applied inconsistently. While the number of companies with Frontier AI Safety Frameworks has more than doubled, and new technical methods have been developed, the overall effectiveness of these measures in protecting against mounting potential harms remains uncertain, and significant disagreement exists about whether current approaches are appropriate or sufficient. This uncertainty reflects the limited availability of shared metrics and evaluation methods to assess the maturity and real-world performance of safeguards over time. Developing such approaches will be essential to track progress and ensure that technical and non-technical risk mitigation measures keep pace with advancing capabilities.

# References

An asterisk (*) denotes that the reference was either published by an AI company or at least 50% of the authors of a preprint have a for-profit AI company as their affiliation.

1  Y. Bengio, S. Clare, C. Prunkl, M. Andriushchenko, B. Bucknall, P. Fox, T. Hu, C. Jones, S. Manning, N. Maslej, V. Mavroudis, C. McGlynn, M. Murray, S. Rismani, C. Stix, L. Velasco, N. Wheeler, … Y.-Q. Zhang, "International AI Safety Report: First Key Update: Capabilities and Risk Implications" (DSIT, 2025); https://internationalaisafetyreport.org/publication/first-key-update-capabilities-and-risk-implications.

2*  OpenAI, "GPT-5 System Card" (OpenAI, 2025); https://cdn.openai.com/gpt-5-system-card.pdf.

3*  Google, "Gemini 2.5 Deep Think - Model Card" (Google, 2025); https://storage.googleapis.com/deepmind-media/Model-Cards/Gemini-2-5-Deep-Think-Model-Card.pdf.

4*  Anthropic, "System Card: Claude Opus 4 & Claude Sonnet 4" (Anthropic, 2025); https://www-cdn.anthropic.com/07b2a3f9902ee19fe39a36ca638e5ae987bc64dd.pdf.

5  Y. Bengio, T. Maharaj, L. Ong, S. Russell, D. Song, M. Tegmark, L. Xue, Y.-Q. Zhang, S. Casper, W. S. Lee, S. Mindermann, V. Wilfred, V. Balachandran, F. Barez, M. Belinsky, I. Bello, M. Bourgon, … D. Žikelić, The Singapore Consensus on Global AI Safety Research Priorities, *arXiv [cs.AI]* (2025); http://arxiv.org/abs/2506.20702.

6  A. Reuel, B. Bucknall, S. Casper, T. Fist, L. Soder, O. Aarne, L. Hammond, L. Ibrahim, A. Chan, P. Wills, M. Anderljung, B. Garfinkel, L. Heim, A. Trask, G. Mukobi, R. Schaeffer, M. Baker, … R. Trager, Open Problems in Technical AI Governance, *arXiv [cs.CY]* (2024); http://arxiv.org/abs/2407.14981.

7  L. Ouyang, J. Wu, X. Jiang, D. Almeida, C. Wainwright, P. Mishkin, C. Zhang, S. Agarwal, K. Slama, A. Gray, J. Schulman, J. Hilton, F. Kelton, L. Miller, M. Simens, A. Askell, P. Welinder, … R. Lowe, "Training Language Models to Follow Instructions with Human Feedback" in *36th Conference on Neural Information Processing Systems (NeurIPS 2022)* (New Orleans, LA, USA, 2022); https://openreview.net/forum?id=TG8KACxEON.

8  A. Zou, L. Phan, J. Wang, D. Duenas, M. Lin, M. Andriushchenko, R. Wang, Z. Kolter, M. Fredrikson, D. Hendrycks, Improving Alignment and Robustness with Circuit Breakers. *Neural Information Processing Systems*, 83345–83373 (2024); https://proceedings.neurips.cc/paper_files/paper/2024/hash/97ca7168c2c333df5ea61ece3b3276e1-Abstract-Conference.html.

9*  M. Sharma, M. Tong, J. Mu, J. Wei, J. Kruthoff, S. Goodfriend, E. Ong, A. Peng, R. Agarwal, C. Anil, A. Askell, N. Bailey, J. Benton, E. Bluemke, S. R. Bowman, E. Christiansen, H. Cunningham, … E. Perez, Constitutional Classifiers: Defending against Universal Jailbreaks across Thousands of Hours of Red Teaming, *arXiv [cs.CL]* (2025); http://arxiv.org/abs/2501.18837.

10  A. Peng, J. Michael, H. Sleight, E. Perez, M. Sharma, Rapid Response: Mitigating LLM Jailbreaks with a Few Examples, *arXiv [cs.CL]* (2024); http://arxiv.org/abs/2411.07494.

11  P. Maini, S. Goyal, D. Sam, A. Robey, Y. Savani, Y. Jiang, A. Zou, Z. C. Lipton, J. Z. Kolter, Safety Pretraining: Toward the next Generation of Safe AI, *arXiv [cs.LG]* (2025); http://arxiv.org/abs/2504.16980.

12*  E. Wallace, O. Watkins, M. Wang, K. Chen, C. Koch, "Estimating Worst-Case Frontier Risks of Open-Weight LLMs" (OpenAI, 2025); https://cdn.openai.com/pdf/231bf018-659a-494d-976c-2efdfc72b652/oai_gpt-oss_Model_Safety.pdf.

13  K. O'Brien, S. Casper, Q. Anthony, T. Korbak, R. Kirk, X. Davies, I. Mishra, G. Irving, Y. Gal, S. Biderman, Deep Ignorance: Filtering Pretraining Data Builds Tamper-Resistant Safeguards into Open-Weight LLMs, *arXiv [cs.LG]* (2025); http://arxiv.org/abs/2508.06601.

14  K. Li, Y. Chen, F. Viégas, M. Wattenberg, When Bad Data Leads to Good Models, *arXiv [cs.LG]* (2025); http://arxiv.org/abs/2505.04741.

**15** A. Paullada, I. D. Raji, E. M. Bender, E. Denton, A. Hanna, Data and Its (dis)contents: A Survey of Dataset Development and Use in Machine Learning Research. *Patterns* **2**, 100336 (2021); https://doi.org/10.1016/j.patter.2021.100336.

**16** L. Berti, F. Giorgi, G. Kasneci, Emergent Abilities in Large Language Models: A Survey, *arXiv [cs.LG]* (2025); http://arxiv.org/abs/2503.05788.

**17** T. Huang, S. Hu, F. Ilhan, S. F. Tekin, Z. Yahn, Y. Xu, L. Liu, Safety Tax: Safety Alignment Makes Your Large Reasoning Models Less Reasonable, *arXiv [cs.CR]* (2025); http://arxiv.org/abs/2503.00555.

**18** P. Peigné, M. Kniejski, F. Sondej, M. David, J. Hoelscher-Obermaier, C. Schroeder de Witt, E. Kran, Multi-Agent Security Tax: Trading off Security and Collaboration Capabilities in Multi-Agent Systems. *Proceedings of the 39th AAAI Conference on Artificial Intelligence* **39**, 27573–27581 (2025); https://doi.org/10.1609/aaai.v39i26.34970.

**19\*** D. M. Ziegler, N. Stiennon, J. Wu, T. B. Brown, A. Radford, D. Amodei, P. Christiano, G. Irving, "Fine-Tuning Language Models from Human Preferences" (OpenAI, 2020); http://arxiv.org/abs/1909.08593.

**20** S. Casper, X. Davies, C. Shi, T. K. Gilbert, J. Scheurer, J. Rando, R. Freedman, T. Korbak, D. Lindner, P. Freire, T. T. Wang, S. Marks, C.-R. Segerie, M. Carroll, A. Peng, P. Christoffersen, M. Damani, … D. Hadfield-Menell, Open Problems and Fundamental Limitations of Reinforcement Learning from Human Feedback. *Transactions on Machine Learning Research* (2023); https://openreview.net/forum?id=bx24KpJ4Eb.

**21** M. Glickman, T. Sharot, How Human-AI Feedback Loops Alter Human Perceptual, Emotional and Social Judgements. *Nature Human Behaviour* **9**, 345–359 (2025); https://doi.org/10.1038/s41562-024-02077-2.

**22** K. Kobalczyk, M. Van Der Schaar, "Preference Learning for AI Alignment: A Causal Perspective" in *Proceedings of the 42nd International Conference on Machine Learning*, A. Singh, M. Fazel, D. Hsu, S. Lacoste-Julien, F. Berkenkamp, T. Maharaj, K. Wagstaff, J. Zhu, Eds. (PMLR, 13–19 Jul 2025) vol. 267 of *Proceedings of Machine Learning Research*, pp. 31063–31083; https://proceedings.mlr.press/v267/kobalczyk25a.html.

**23\*** X. Wen, J. Lou, X. Lu, J. Yang, Y. Liu, Y. Lu, D. Zhang, X. Yu, Scalable Oversight for Superhuman AI via Recursive Self-Critiquing, *arXiv [cs.AI]* (2025); http://arxiv.org/abs/2502.04675.

**24** D. Dai, M. Liu, A. Li, J. Cao, Y. Wang, C. Wang, X. Peng, Z. Zheng, FeedbackEval: A Benchmark for Evaluating Large Language Models in Feedback-Driven Code Repair Tasks, *arXiv [cs.SE]* (2025); http://arxiv.org/abs/2504.06939.

**25\*** Z. Wang, J. Zeng, O. Delalleau, H.-C. Shin, F. Soares, A. Bukharin, E. Evans, Y. Dong, O. Kuchaiev, HelpSteer3-Preference: Open Human-Annotated Preference Data across Diverse Tasks and Languages, *arXiv [cs.CL]* (2025); http://arxiv.org/abs/2505.11475.

**26** N. Lambert, J. Morrison, V. Pyatkin, S. Huang, H. Ivison, F. Brahman, L. J. V. Miranda, A. Liu, N. Dziri, X. Lyu, Y. Gu, S. Malik, V. Graf, J. D. Hwang, J. Yang, R. Le Bras, O. Tafjord, … H. Hajishirzi, "Tulu 3: Pushing Frontiers in Open Language Model Post-Training" in *Second Conference on Language Modeling* (2025); https://openreview.net/forum?id=i1uGbfHHpH#discussion.

**27** M. Mazeika, L. Phan, X. Yin, A. Zou, Z. Wang, N. Mu, E. Sakhaee, N. Li, S. Basart, B. Li, D. Forsyth, D. Hendrycks, HarmBench: A Standardized Evaluation Framework for Automated Red Teaming and Robust Refusal, *arXiv [cs.LG]* (2024); http://arxiv.org/abs/2402.04249.

**28** S. Lee, M. Kim, L. Cherif, D. Dobre, J. Lee, S. J. Hwang, K. Kawaguchi, G. Gidel, Y. Bengio, N. Malkin, M. Jain, Learning Diverse Attacks on Large Language Models for Robust Red-Teaming and Safety Tuning, *arXiv [cs.CL]* (2024); http://arxiv.org/abs/2405.18540.

**29\*** N. Howe, I. McKenzie, O. Hollinsworth, M. Zajac, T. Tseng, A. Tucker, P.-L. Bacon, A. Gleave, Scaling Trends in Language Model Robustness, *arXiv [cs.LG]* (2024); http://arxiv.org/abs/2407.18213.

**30\*** A. Cuevas, S. Dash, B. K. Nayak, D. Vann, M. I. G. Daepp, Anecdoctoring: Automated Red-Teaming across Language and Place, *arXiv [cs.CL]* (2025); http://arxiv.org/abs/2509.19143.

**31** Z.-W. Hong, I. Shenfeld, T.-H. Wang, Y.-S. Chuang, A. Pareja, J. Glass, A. Srivastava, P. Agrawal, Curiosity-Driven Red-Teaming for Large Language Models, *arXiv [cs.LG]* (2024); http://arxiv.org/abs/2402.19464.

**32** T. Yun, P.-L. St-Charles, J. Park, Y. Bengio, M. Kim, Active Attacks: Red-Teaming LLMs via Adaptive Environments, *arXiv [cs.LG]* (2025); http://arxiv.org/abs/2509.21947.

**33** S. Xhonneux, A. Sordoni, S. Günnemann, G. Gidel, L. Schwinn, "Efficient Adversarial Training in LLMs with Continuous Attacks" in *38th Annual Conference on Neural Information Processing Systems (NeurIPS)* (2024); https://openreview.net/pdf?id=8jB6sGqvgQ.

**34** S. Casper, L. Schulze, O. Patel, D. Hadfield-Menell, Defending Against Unforeseen Failure Modes with Latent Adversarial Training, *arXiv [cs.CR]* (2024); http://dx.doi.org/10.48550/arXiv.2403.05030.

**35** A. Sheshadri, A. Ewart, P. Guo, A. Lynch, C. Wu, V. Hebbar, H. Sleight, A. C. Stickland, E. Perez, D. Hadfield-Menell, S. Casper, Latent Adversarial Training Improves Robustness to Persistent Harmful Behaviors in LLMs, *arXiv [cs.LG]* (2024); http://arxiv.org/abs/2407.15549.

**36** C. Dékány, S. Balauca, R. Staab, D. I. Dimitrov, M. Vechev, MixAT: Combining Continuous and Discrete Adversarial Training for LLMs, *arXiv [cs.LG]* (2025); http://arxiv.org/abs/2505.16947.

**37** M. Andriushchenko, F. Croce, N. Flammarion, Jailbreaking Leading Safety-Aligned LLMs with Simple Adaptive Attacks, *arXiv [cs.CR]* (2024); http://arxiv.org/abs/2404.02151.

**38*** N. Li, Z. Han, I. Steneker, W. Primack, R. Goodside, H. Zhang, Z. Wang, C. Menghini, S. Yue, LLM Defenses Are Not Robust to Multi-Turn Human Jailbreaks yet, *arXiv [cs.LG]* (2024); http://arxiv.org/abs/2408.15221.

**39*** A. Zou, M. Lin, E. Jones, M. Nowak, M. Dziemian, N. Winter, A. Grattan, V. Nathanael, A. Croft, X. Davies, J. Patel, R. Kirk, N. Burnikell, Y. Gal, D. Hendrycks, J. Z. Kolter, M. Fredrikson, Security Challenges in AI Agent Deployment: Insights from a Large Scale Public Competition, *arXiv [cs.AI]* (2025); http://arxiv.org/abs/2507.20526.

**40** D. Lüdke, T. Wollschläger, P. Ungermann, S. Günnemann, L. Schwinn, Diffusion LLMs Are Natural Adversaries for Any LLM, *arXiv [cs.LG]* (2025); http://arxiv.org/abs/2511.00203.

**41** Anthropic, "System Card: Claude Sonnet 4.5" (Anthropic, 2025); https://assets.anthropic.com/m/12f214efcc2f457a/original/Claude-Sonnet-4-5-System-Card.pdf.

**42** A. Souly, J. Rando, E. Chapman, X. Davies, B. Hasircioglu, E. Shereen, C. Mougan, V. Mavroudis, E. Jones, C. Hicks, N. Carlini, Y. Gal, R. Kirk, Poisoning Attacks on LLMs Require a near-Constant Number of Poison Samples, *arXiv [cs.LG]* (2025); http://arxiv.org/abs/2510.07192.

**43*** M. Wang, T. D. la Tour, O. Watkins, A. Makelov, R. A. Chi, S. Miserendino, J. Wang, A. Rajaram, J. Heidecke, T. Patwardhan, D. Mossing, Persona Features Control Emergent Misalignment, *arXiv [cs.LG]* (2025); http://arxiv.org/abs/2506.19823.

**44** J. Betley, D. C. H. Tan, N. Warncke, A. Sztyber-Betley, X. Bao, M. Soto, N. Labenz, O. Evans, "Emergent Misalignment: Narrow Finetuning Can Produce Broadly Misaligned LLMs" in *Forty-Second International Conference on Machine Learning* (2025); https://openreview.net/forum?id=aOIJ2gVRWW.

**45** Z. Zhang, S. Cui, Y. Lu, J. Zhou, J. Yang, H. Wang, M. Huang, Agent-SafetyBench: Evaluating the Safety of LLM Agents, *arXiv [cs.CL]* (2024); http://arxiv.org/abs/2412.14470.

**46** X. Li, R. Wang, M. Cheng, T. Zhou, C.-J. Hsieh, "DrAttack: Prompt Decomposition and Reconstruction Makes Powerful LLMs Jailbreakers" in *Findings of the Association for Computational Linguistics: EMNLP 2024* (Association for Computational Linguistics, Stroudsburg, PA, USA, 2024), pp. 13891–13913; https://doi.org/10.18653/v1/2024.findings-emnlp.813.

**47** M. Andriushchenko, A. Souly, M. Dziemian, D. Duenas, M. Lin, J. Wang, D. Hendrycks, A. Zou, Z. Kolter, M. Fredrikson, E. Winsor, J. Wynne, Y. Gal, X. Davies, AgentHarm: A Benchmark for Measuring Harmfulness of LLM Agents, *arXiv [cs.LG]* (2024); http://arxiv.org/abs/2410.09024.

**48** T. Kuntz, A. Duzan, H. Zhao, F. Croce, Z. Kolter, N. Flammarion, M. Andriushchenko, OS-Harm: A Benchmark for Measuring Safety of Computer

Use Agents, *arXiv [cs.SE]* (2025); http://arxiv.org/abs/2506.14866.

49  A. Naik, P. Quinn, G. Bosch, E. Gouné, F. J. C. Zabala, J. R. Brown, E. J. Young, AgentMisalignment: Measuring the Propensity for Misaligned Behaviour in LLM-Based Agents, *arXiv [cs.AI]* (2025); http://arxiv.org/abs/2506.04018.

50  C. Yu, B. Stroebl, D. Yang, O. Papakyriakopoulos, Safety Devolution in AI Agents, *arXiv [cs.CY]* (2025); http://arxiv.org/abs/2505.14215.

51  J. Y. F. Chiang, S. Lee, J.-B. Huang, F. Huang, Y. Chen, Why Are Web AI Agents More Vulnerable than Standalone LLMs? A Security Analysis, *arXiv [cs.LG]* (2025); http://arxiv.org/abs/2502.20383.

52  B. Cottier, J. You, N. Martemianova, D. Owen, "How Far Behind Are Open Models?" (Epoch AI, 2024); https://epoch.ai/blog/open-models-report.

53  N. Maslej, L. Fattorini, R. Perrault, Y. Gil, V. Parli, N. Kariuki, E. Capstick, A. Reuel, E. Brynjolfsson, J. Etchemendy, K. Ligett, T. Lyons, J. Manyika, J. C. Niebles, Y. Shoham, R. Wald, T. Walsh, … S. Oak, "The AI Index 2025 Annual Report" (AI Index Steering Committee, Institute for Human-Centered AI, Stanford University, 2025); https://hai.stanford.edu/assets/files/hai_ai_index_report_2025.pdf.

54*  A. Yang, A. Li, B. Yang, B. Zhang, B. Hui, B. Zheng, B. Yu, C. Gao, C. Huang, C. Lv, C. Zheng, D. Liu, F. Zhou, F. Huang, F. Hu, H. Ge, H. Wei, … Z. Qiu, Qwen3 Technical Report, *arXiv [cs.CL]* (2025); http://arxiv.org/abs/2505.09388.

55*  Meta AI, The Llama 4 Herd: The Beginning of a New Era of Natively Multimodal AI Innovation, *Meta AI* (2025); https://ai.meta.com/blog/llama-4-multimodal-intelligence/.

56*  Kimi Team, Y. Bai, Y. Bao, G. Chen, J. Chen, N. Chen, R. Chen, Y. Chen, Y. Chen, Y. Chen, Z. Chen, J. Cui, H. Ding, M. Dong, A. Du, C. Du, D. Du, … X. Zu, Kimi K2: Open Agentic Intelligence, *arXiv [cs.LG]* (2025); http://arxiv.org/abs/2507.20534.

57*  OpenAI, S. Agarwal, L. Ahmad, J. Ai, S. Altman, A. Applebaum, E. Arbus, R. K. Arora, Y. Bai, B. Baker, H. Bao, B. Barak, A. Bennett, T. Bertao, N. Brett, E. Brevdo, G. Brockman, … S. Zhao, Gpt-Oss-120b & Gpt-Oss-20b Model Card, *arXiv [cs.CL]* (2025); https://cdn.openai.com/pdf/419b6906-9da6-406c-a19d-1bb078ac7637/oai_gpt-oss_model_card.pdf.

58*  GLM-4.5 Team, A. Zeng, X. Lv, Q. Zheng, Z. Hou, B. Chen, C. Xie, C. Wang, D. Yin, H. Zeng, J. Zhang, K. Wang, L. Zhong, M. Liu, R. Lu, S. Cao, X. Zhang, … J. Tang, GLM-4.5: Agentic, Reasoning, and Coding (ARC) Foundation Models, *arXiv [cs.CL]* (2025); http://arxiv.org/abs/2508.06471.

59  D. Guo, D. Yang, H. Zhang, J. Song, P. Wang, Q. Zhu, R. Xu, R. Zhang, S. Ma, X. Bi, X. Zhang, X. Yu, Y. Wu, Z. F. Wu, Z. Gou, Z. Shao, Z. Li, … Z. Zhang, DeepSeek-R1 Incentivizes Reasoning in LLMs through Reinforcement Learning. *Nature* **645**, 633–638 (2025); https://doi.org/10.1038/s41586-025-09422-z.

60*  C. Wu, J. Li, J. Zhou, J. Lin, K. Gao, K. Yan, S.-M. Yin, S. Bai, X. Xu, Y. Chen, Y. Chen, Z. Tang, Z. Zhang, Z. Wang, A. Yang, B. Yu, C. Cheng, … Z. Liu, Qwen-Image Technical Report, *arXiv [cs.CV]* (2025); http://arxiv.org/abs/2508.02324.

61*  S. Cao, H. Chen, P. Chen, Y. Cheng, Y. Cui, X. Deng, Y. Dong, K. Gong, T. Gu, X. Gu, T. Hang, D. Huang, J. Jiang, Z. Jiang, W. Kong, C. Li, D. Li, … Z. Zhong, HunyuanImage 3.0 Technical Report, *arXiv [cs.CV]* (2025); http://dx.doi.org/10.48550/arXiv.2509.23951.

62*  T. Wan, A. Wang, B. Ai, B. Wen, C. Mao, C.-W. Xie, D. Chen, F. Yu, H. Zhao, J. Yang, J. Zeng, J. Wang, J. Zhang, J. Zhou, J. Wang, J. Chen, K. Zhu, … Z. Liu, Wan: Open and Advanced Large-Scale Video Generative Models, *arXiv [cs.CV]* (2025); http://dx.doi.org/10.48550/arXiv.2503.20314.

63*  W. Kong, Q. Tian, Z. Zhang, R. Min, Z. Dai, J. Zhou, J. Xiong, X. Li, B. Wu, J. Zhang, K. Wu, Q. Lin, J. Yuan, Y. Long, A. Wang, A. Wang, C. Li, … C. Zhong, HunyuanVideo: A Systematic Framework for Large Video Generative Models, *arXiv [cs.CV]* (2024); http://dx.doi.org/10.48550/arXiv.2412.03603.

64  R. Bommasani, S. Kapoor, K. Klyman, S. Longpre, A. Ramaswami, D. Zhang, M. Schaake, D. E. Ho, A. Narayanan, P. Liang, Considerations for Governing Open Foundation Models. *Science* **386**, 151–153 (2024); https://doi.org/10.1126/science.adp1848.

65  US National Telecommunications and Information Administration, "Dual-Use Foundation Models with Widely Available Model Weights NTIA

Report" (US Department of Commerce, 2024); https://www.ntia.gov/issues/artificial-intelligence/open-model-weights-report.

66* E. Seger, B. O'Dell, "Open Horizons: Exploring Nuanced Technical and Policy Approaches to Openness in AI" (Demos and Mozilla, 2024); https://demos.co.uk/wp-content/uploads/2024/08/Mozilla-Report_2024.pdf.

67 F. Eiras, A. Petrov, B. Vidgen, C. Schroeder, F. Pizzati, K. Elkins, S. Mukhopadhyay, A. Bibi, A. Purewal, C. Botos, F. Steibel, F. Keshtkar, F. Barez, G. Smith, G. Guadagni, J. Chun, J. Cabot, … J. Foerster, Risks and Opportunities of Open-Source Generative AI, *arXiv [cs.LG]* (2024); http://arxiv.org/abs/2405.08597.

68 S. Kapoor, R. Bommasani, K. Klyman, S. Longpre, A. Ramaswami, P. Cihon, A. Hopkins, K. Bankston, S. Biderman, M. Bogen, R. Chowdhury, A. Engler, P. Henderson, Y. Jernite, S. Lazar, S. Maffulli, A. Nelson, … A. Narayanan, On the Societal Impact of Open Foundation Models, *arXiv [cs.CY]* (2024); http://arxiv.org/abs/2403.07918.

69 C. François, L. Péran, A. Bdeir, N. Dziri, W. Hawkins, Y. Jernite, S. Kapoor, J. Shen, H. Khlaaf, K. Klyman, N. Marda, M. Pellat, D. Raji, D. Siddarth, A. Skowron, J. Spisak, M. Srikumar, … J. Weedon, A Different Approach to AI Safety: Proceedings from the Columbia Convening on Openness in Artificial Intelligence and AI Safety, *arXiv [cs.AI]* (2025); http://dx.doi.org/10.48550/arXiv.2506.22183.

70 Y. Bengio, S. Mindermann, D. Privitera, T. Besiroglu, R. Bommasani, S. Casper, Y. Choi, P. Fox, B. Garfinkel, D. Goldfarb, H. Heidari, A. Ho, S. Kapoor, L. Khalatbari, S. Longpre, S. Manning, V. Mavroudis, … Y. Zeng, "International AI Safety Report" (Department for Science, Innovation and Technology, 2025); https://www.gov.uk/government/publications/international-ai-safety-report-2025.

71 E. Seger, N. Dreksler, R. Moulange, E. Dardaman, J. Schuett, K. Wei, C. Winter, M. Arnold, S. Ó. hÉigeartaigh, A. Korinek, M. Anderljung, B. Bucknall, A. Chan, E. Stafford, L. Koessler, A. Ovadya, B. Garfinkel, … A. Gupta, "Open-Sourcing Highly Capable Foundation Models: An Evaluation of Risks, Benefits, and Alternative Methods for Pursuing Open-Source Objectives" ( Centre for the Governance of AI, 2023); http://arxiv.org/abs/2311.09227.

72 A. Chan, B. Bucknall, H. Bradley, D. Krueger, Hazards from Increasingly Accessible Fine-Tuning of Downloadable Foundation Models, *arXiv [cs.LG]* (2023); http://arxiv.org/abs/2312.14751.

73 T. Huang, S. Hu, F. Ilhan, S. F. Tekin, L. Liu, Harmful Fine-Tuning Attacks and Defenses for Large Language Models: A Survey, *arXiv [cs.CR]* (2024); http://arxiv.org/abs/2409.18169.

74 IWF, "What Has Changed in the AI CSAM Landscape?" (Internet Watch Foundation, 2024).

75 W. Hawkins, C. Russell, B. Mittelstadt, Deepfakes on Demand: The Rise of Accessible Non-Consensual Deepfake Image Generators, *arXiv [cs.CY]* (2025); http://arxiv.org/abs/2505.03859.

76 E. H. Vaughan, NCMEC Releases New Data: 2024 in Numbers, *National Center for Missing & Exploited Children* (2025); http://www.ncmec.org/content/ncmec/en/blog/2025/ncmec-releases-new-data-2024-in-numbers.html.

77 S. Casper, K. O'Brien, S. Longpre, E. Seger, K. Klyman, R. Bommasani, A. Nrusimha, I. Shumailov, S. Mindermann, S. Basart, F. Rudzicz, K. Pelrine, A. Ghosh, A. Strait, R. Kirk, D. Hendrycks, P. Henderson, … D. Hadfield-Menell, Open Technical Problems in Open-Weight AI Model Risk Management, *Social Science Research Network* (2025); https://doi.org/10.2139/ssrn.5705186.

78 M. Srikumar, J. Chang, K. Chmielinski, "Risk Mitigation Strategies for the Open Foundation Model Value Chain: Insights from PAI Workshop Co-Hosted with GitHub" (Partnership on AI, 2024); https://partnershiponai.notion.site/1e8a6131dda045f1ad00054933b0bda0?v=dcb890146f7d464a86f11fcd5de372c0.

79 AI Security Institute, Managing Risks from Increasingly Capable Open-Weight AI Systems, *AI Security Institute* (2025); https://www.aisi.gov.uk/work/managing-risks-from-increasingly-capable-open-weight-ai-systems.

80 P. Henderson, E. Mitchell, C. Manning, D. Jurafsky, C. Finn, "Self-Destructing Models: Increasing the Costs of Harmful Dual Uses of Foundation Models" in *Proceedings of the 2023 AAAI/ACM Conference on AI, Ethics, and Society* (Association for Computing Machinery, New York, NY, USA, 2023), *AIES '23*, pp. 287–296; https://doi.org/10.1145/3600211.3604690.

**81*** C. Fan, J. Jia, Y. Zhang, A. Ramakrishna, M. Hong, S. Liu, Towards LLM Unlearning Resilient to Relearning Attacks: A Sharpness-Aware Minimization Perspective and beyond, *arXiv [cs.LG]* (2025); http://arxiv.org/abs/2502.05374.

**82** D. Rosati, J. Wehner, K. Williams, Ł. Bartoszcze, D. Atanasov, R. Gonzales, S. Majumdar, C. Maple, H. Sajjad, F. Rudzicz, Representation Noising Effectively Prevents Harmful Fine-Tuning on LLMs, *arXiv [cs.CL]* (2024); http://arxiv.org/abs/2405.14577.

**83** R. Tamirisa, B. Bharathi, L. Phan, A. Zhou, A. Gatti, T. Suresh, M. Lin, J. Wang, R. Wang, R. Arel, A. Zou, D. Song, B. Li, D. Hendrycks, M. Mazeika, Tamper-Resistant Safeguards for Open-Weight LLMs, *arXiv [cs.LG]* (2024); http://arxiv.org/abs/2408.00761.

**84** A. Abdalla, I. Shaheen, D. DeGenaro, R. Mallick, B. Raita, S. A. Bargal, GIFT: Gradient-Aware Immunization of Diffusion Models against Malicious Fine-Tuning with Safe Concepts Retention, *arXiv [cs.CR]* (2025); http://arxiv.org/abs/2507.13598.

**85*** A. F. Cooper, C. A. Choquette-Choo, M. Bogen, M. Jagielski, K. Filippova, K. Z. Liu, A. Chouldechova, J. Hayes, Y. Huang, N. Mireshghallah, I. Shumailov, E. Triantafillou, P. Kairouz, N. Mitchell, P. Liang, D. E. Ho, Y. Choi, … K. Lee, Machine Unlearning Doesn't Do What You Think: Lessons for Generative AI Policy, Research, and Practice, *arXiv [cs.LG]* (2024); http://arxiv.org/abs/2412.06966.

**86** J. Łucki, B. Wei, Y. Huang, P. Henderson, F. Tramèr, J. Rando, An Adversarial Perspective on Machine Unlearning for AI Safety, *arXiv [cs.LG]* (2024); http://arxiv.org/abs/2409.18025.

**87** S. Hu, Y. Fu, Z. S. Wu, V. Smith, Jogging the Memory of Unlearned LLMs through Targeted Relearning Attacks, *arXiv [cs.LG]* (2024); http://arxiv.org/abs/2406.13356.

**88*** A. Deeb, F. Roger, Do Unlearning Methods Remove Information from Language Model Weights?, *arXiv [cs.LG]* (2024); http://arxiv.org/abs/2410.08827.

**89** Z. Che, S. Casper, R. Kirk, A. Satheesh, S. Slocum, L. E. McKinney, R. Gandikota, A. Ewart, D. Rosati, Z. Wu, Z. Cai, B. Chughtai, Y. Gal, F. Huang, D. Hadfield-Menell, Model Tampering Attacks Enable More Rigorous Evaluations of LLM Capabilities, *arXiv [cs.CR]* (2025); http://arxiv.org/abs/2502.05209.

**90** J. Luo, T. Ding, K. H. R. Chan, D. Thaker, A. Chattopadhyay, C. Callison-Burch, R. Vidal, PaCE: Parsimonious Concept Engineering for Large Language Models, *arXiv [cs.CL]* (2024); http://arxiv.org/abs/2406.04331.

**91** D. Gottesman, M. Geva, "Estimating Knowledge in Large Language Models without Generating a Single Token" in *Proceedings of the 2024 Conference on Empirical Methods in Natural Language Processing* (Association for Computational Linguistics, Stroudsburg, PA, USA, 2024), pp. 3994–4019; https://doi.org/10.18653/v1/2024.emnlp-main.232.

**92** O. Aarne, T. Fist, C. Withers, "Secure, Governable Chips: Using On-Chip Mechanisms to Manage National Security Risks from AI & Advanced Computing" (Center for a New American Security, 2024); https://s3.us-east-1.amazonaws.com/files.cnas.org/documents/CNAS-Report-Tech-Secure-Chips-Jan-24-finalb.pdf.

**93** A. O'Gara, G. Kulp, W. Hodgkins, J. Petrie, V. Immler, A. Aysu, K. Basu, S. Bhasin, S. Picek, A. Srivastava, Hardware-Enabled Mechanisms for Verifying Responsible AI Development, *arXiv [cs.CR]* (2025); http://arxiv.org/abs/2505.03742.

**94** C. Yueh-Han, N. Joshi, Y. Chen, M. Andriushchenko, R. Angell, H. He, Monitoring Decomposition Attacks in LLMs with Lightweight Sequential Monitors, *arXiv [cs.CR]* (2025); http://arxiv.org/abs/2506.10949.

**95** D. Brown, M. Sabbaghi, L. Sun, A. Robey, G. J. Pappas, E. Wong, H. Hassani, Benchmarking Misuse Mitigation against Covert Adversaries, *arXiv [cs.CR]* (2025); http://arxiv.org/abs/2506.06414.

**96*** I. R. McKenzie, O. J. Hollinsworth, T. Tseng, X. Davies, S. Casper, A. D. Tucker, R. Kirk, A. Gleave, STACK: Adversarial Attacks on LLM Safeguard Pipelines, *arXiv [cs.CL]* (2025); http://arxiv.org/abs/2506.24068.

**97** N. Kirch, C. Weisser, S. Field, H. Yannakoudakis, S. Casper, What Features in Prompts Jailbreak LLMs? Investigating the Mechanisms behind Attacks, *arXiv [cs.CR]* (2024); https://doi.org/10.48550/ARXIV.2411.03343.

98  N. Goldowsky-Dill, B. Chughtai, S. Heimersheim, M. Hobbhahn, Detecting Strategic Deception Using Linear Probes, *arXiv [cs.LG]* (2025); http://arxiv.org/abs/2502.03407.

99*  B. Baker, J. Huizinga, L. Gao, Z. Dou, M. Y. Guan, A. Madry, W. Zaremba, J. Pachocki, D. Farhi, "Monitoring Reasoning Models for Misbehavior and the Risks of Promoting Obfuscation" (OpenAI, 2025); https://arxiv.org/abs/2503.11926.

100*  T. Korbak, M. Balesni, E. Barnes, Y. Bengio, J. Benton, J. Bloom, M. Chen, A. Cooney, A. Dafoe, A. Dragan, S. Emmons, O. Evans, D. Farhi, R. Greenblatt, D. Hendrycks, M. Hobbhahn, E. Hubinger, … V. Mikulik, Chain of Thought Monitorability: A New and Fragile Opportunity for AI Safety, *arXiv [cs.AI]* (2025); http://arxiv.org/abs/2507.11473.

101  R. Greenblatt, B. Shlegeris, K. Sachan, F. Roger, AI Control: Improving Safety Despite Intentional Subversion, *arXiv [cs.LG]* (2023); http://dx.doi.org/10.48550/arXiv.2312.06942.

102  Y. Bengio, M. K. Cohen, N. Malkin, M. MacDermott, D. Fornasiere, P. Greiner, Y. Kaddar, Can a Bayesian Oracle Prevent Harm from an Agent?, *arXiv [cs.AI]* (2024); http://arxiv.org/abs/2408.05284.

103  T. Korbak, J. Clymer, B. Hilton, B. Shlegeris, G. Irving, A Sketch of an AI Control Safety Case, *arXiv [cs.AI]* (2025); http://arxiv.org/abs/2501.17315.

104  V. Kovarik, E. O. Chen, S. Petersen, A. Ghersengorin, V. Conitzer, AI Testing Should Account for Sophisticated Strategic Behaviour, *arXiv [cs.GT]* (2025); http://arxiv.org/abs/2508.14927.

105*  OpenAI, "Operator System Card" (OpenAI, 2025); https://cdn.openai.com/operator_system_card.pdf.

106  L. Zhu, Q. Lu, D. Ming, S. U. Lee, C. Wang, Designing Meaningful Human Oversight in AI, *Social Science Research Network* (2025); https://doi.org/10.2139/ssrn.5501939.

107*  H. Mozannar, G. Bansal, C. Tan, A. Fourney, V. Dibia, J. Chen, J. Gerrits, T. Payne, M. K. Maldaner, M. Grunde-McLaughlin, E. Zhu, G. Bassman, J. Alber, P. Chang, R. Loynd, F. Niedtner, E. Kamar, … S. Amershi, Magentic-UI: Towards Human-in-the-Loop Agentic Systems, *arXiv [cs.AI]* (2025); http://arxiv.org/abs/2507.22358.

108  T. Hua, J. Baskerville, H. Lemoine, M. Hopman, A. Bhatt, T. Tracy, "Combining Cost Constrained Runtime Monitors for AI Safety" in *The 39th Annual Conference on Neural Information Processing Systems* (2025); https://openreview.net/forum?id=hVR3023UP2.

109  A. McKenzie, U. Pawar, P. Blandfort, W. Bankes, D. Krueger, E. S. Lubana, D. Krasheninnikov, "Detecting High-Stakes Interactions with Activation Probes" in *The 39th Annual Conference on Neural Information Processing Systems* (2025); https://openreview.net/forum?id=8YniJnJQ0P.

110*  OpenAI, "GPT-4o System Card" (OpenAI, 2024); https://cdn.openai.com/gpt-4o-system-card.pdf.

111*  OpenAI, "ChatGPT Agent System Card" (2025); https://cdn.openai.com/pdf/839e66fc-602c-48bf-81d3-b21eacc3459d/chatgpt_agent_system_card.pdf.

112  Organisation for Economic Co-Operation and Development, "Towards a Common Reporting Framework for AI Incidents" (OECD, 2025); https://doi.org/10.1787/f326d4ac-en.

113  N. Yu, V. Skripniuk, S. Abdelnabi, M. Fritz, "Artificial Fingerprinting for Generative Models: Rooting Deepfake Attribution in Training Data" in *2021 IEEE/CVF International Conference on Computer Vision (ICCV)* (IEEE, 2021); https://doi.org/10.1109/iccv48922.2021.01418.

114  F. Boenisch, A Systematic Review on Model Watermarking for Neural Networks. *Frontiers in Big Data* **4** (2021); https://www.frontiersin.org/articles/10.3389/fdata.2021.729663/full.

115  P. Fernandez, G. Couairon, H. Jégou, M. Douze, T. Furon, "The Stable Signature: Rooting Watermarks in Latent Diffusion Models" in *2023 IEEE/CVF International Conference on Computer Vision (ICCV)* (2023), pp. 22409–22420; https://doi.org/10.1109/ICCV51070.2023.02053.

116  M. Christ, S. Gunn, T. Malkin, M. Raykova, Provably Robust Watermarks for Open-Source Language Models, *arXiv [cs.CR]* (2024); http://arxiv.org/abs/2410.18861.

**117\***  X. Xu, Y. Yao, Y. Liu, Learning to Watermark LLM-Generated Text via Reinforcement Learning, *arXiv [cs.LG]* (2024); http://arxiv.org/abs/2403.10553.

**118**  G. Pagnotta, D. Hitaj, B. Hitaj, F. Perez-Cruz, L. V. Mancini, TATTOOED: A Robust Deep Neural Network Watermarking Scheme Based on Spread-Spectrum Channel Coding, *arXiv [cs.CR]* (2022); http://arxiv.org/abs/2202.06091.

**119**  P. Lv, P. Li, S. Zhang, K. Chen, R. Liang, H. Ma, Y. Zhao, Y. Li, A Robustness-Assured White-Box Watermark in Neural Networks. *IEEE Transactions on Dependable and Secure Computing* **20**, 5214–5229 (2023); https://doi.org/10.1109/tdsc.2023.3242737.

**120**  L. Li, B. Jiang, P. Wang, K. Ren, H. Yan, X. Qiu, "Watermarking LLMs with Weight Quantization" in *Findings of the Association for Computational Linguistics: EMNLP 2023*, H. Bouamor, J. Pino, K. Bali, Eds. (Association for Computational Linguistics, Singapore, 2023), pp. 3368–3378; https://doi.org/10.18653/v1/2023.findings-emnlp.220.

**121\***  A. Block, A. Sekhari, A. Rakhlin, GaussMark: A Practical Approach for Structural Watermarking of Language Models, *arXiv [cs.CR]* (2025); http://arxiv.org/abs/2501.13941.

**122**  T. Gloaguen, N. Jovanović, R. Staab, M. Vechev, Towards Watermarking of Open-Source LLMs, *arXiv [cs.CR]* (2025); http://arxiv.org/abs/2502.10525.

**123**  S. Zhu, A. Ahmed, R. Kuditipudi, P. Liang, Independence Tests for Language Models, *arXiv [cs.LG]* (2025); http://arxiv.org/abs/2502.12292.

**124**  E. Horwitz, A. Shul, Y. Hoshen, Unsupervised Model Tree Heritage Recovery, *arXiv [cs.LG]* (2024); http://arxiv.org/abs/2405.18432.

**125**  E. Horwitz, N. Kurer, J. Kahana, L. Amar, Y. Hoshen, We Should Chart an Atlas of All the World's Models, *arXiv [cs.LG]* (2025); http://arxiv.org/abs/2503.10633.

**126**  Organisation for Economic Co-Operation and Development, "Sharing Trustworthy AI Models with Privacy-Enhancing Technologies" (OECD, 2025); https://doi.org/10.1787/a266160b-en.

**127**  S. Chappidi, J. Cobbe, C. Norval, A. Mazumder, J. Singh, Accountability Capture: How Record-Keeping to Support AI Transparency and Accountability (re)shapes Algorithmic Oversight. *Proceedings of the AAAI/ACM Conference on AI, Ethics, and Society* **8**, 554–566 (2025); https://doi.org/10.1609/aies.v8i1.36570.

**128**  X. Zhao, S. Gunn, M. Christ, J. Fairoze, A. Fabrega, N. Carlini, S. Garg, S. Hong, M. Nasr, F. Tramer, S. Jha, L. Li, Y.-X. Wang, D. Song, SoK: Watermarking for AI-Generated Content, *arXiv [cs.CR]* (2024); http://arxiv.org/abs/2411.18479.

**129\***  L. Cao, Watermarking for AI Content Detection: A Review on Text, Visual, and Audio Modalities, *arXiv [cs.CR]* (2025); http://arxiv.org/abs/2504.03765.

**130**  S. Dathathri, A. See, S. Ghaisas, P.-S. Huang, R. McAdam, J. Welbl, V. Bachani, A. Kaskasoli, R. Stanforth, T. Matejovicova, J. Hayes, N. Vyas, M. A. Merey, J. Brown-Cohen, R. Bunel, B. Balle, T. Cemgil, … P. Kohli, Scalable Watermarking for Identifying Large Language Model Outputs. *Nature* **634**, 818–823 (2024); https://doi.org/10.1038/s41586-024-08025-4.

**131**  A. Liu, L. Pan, Y. Lu, J. Li, X. Hu, X. Zhang, L. Wen, I. King, H. Xiong, P. Yu, A Survey of Text Watermarking in the Era of Large Language Models. *ACM Computing Surveys* **57**, 1–36 (2025); https://doi.org/10.1145/3691626.

**132**  Z. Yang, G. Zhao, H. Wu, Watermarking for Large Language Models: A Survey. *Mathematics* **13**, 1420 (2025); https://doi.org/10.3390/math13091420.

**133**  W. Wan, J. Wang, Y. Zhang, J. Li, H. Yu, J. Sun, A Comprehensive Survey on Robust Image Watermarking. *Neurocomputing* **488**, 226–247 (2022); https://doi.org/10.1016/j.neucom.2022.02.083.

**134**  M. S. Uddin, Ohidujjaman, M. Hasan, T. Shimamura, Audio Watermarking: A Comprehensive Review. *International Journal of Advanced Computer Science and Applications* **15** (2024); https://doi.org/10.14569/IJACSA.2024.01505141.

**135**  S. Singhi, A. Yadav, A. Gupta, S. Ebrahimi, P. Hassanizadeh, Provenance Detection for AI-Generated Images: Combining Perceptual Hashing, Homomorphic Encryption, and AI Detection Models, *arXiv [cs.CV]* (2025); http://arxiv.org/abs/2503.11195.

# References

**136**  R. Chen, Y. Wu, J. Guo, H. Huang, Improved Unbiased Watermark for Large Language Models, *arXiv [cs.CL]* (2025); http://arxiv.org/abs/2502.11268.

**137**  C2PA Technical Working Group, "C2PA Content Credentials Explained: Addressing Common Questions and Updates" (C2PA, 2025); https://c2pa.org/wp-content/uploads/sites/33/2025/10/content_credentials_wp_0925.pdf.

**138**  A. Knott, D. Pedreschi, R. Chatila, T. Chakraborti, S. Leavy, R. Baeza-Yates, D. Eyers, A. Trotman, P. D. Teal, P. Biecek, S. Russell, Y. Bengio, Generative AI Models Should Include Detection Mechanisms as a Condition for Public Release. *Ethics and Information Technology* **25**, 55 (2023); https://doi.org/10.1007/s10676-023-09728-4.

**139**  L. Lin, N. Gupta, Y. Zhang, H. Ren, C.-H. Liu, F. Ding, X. Wang, X. Li, L. Verdoliva, S. Hu, Detecting Multimedia Generated by Large AI Models: A Survey, *arXiv [cs.MM]* (2024); https://www.techrxiv.org/users/723084/articles/707949-detecting-multimedia-generated-by-large-ai-models-a-survey?commit=17e92ea8d954e6c448a006d4f4e7fd594c9f6f0d.

**140**  A. Hans, A. Schwarzschild, V. Cherepanova, H. Kazemi, A. Saha, M. Goldblum, J. Geiping, T. Goldstein, Spotting LLMs with Binoculars: Zero-Shot Detection of Machine-Generated Text, *arXiv [cs.CL]* (2024); http://arxiv.org/abs/2401.12070.

**141\***  V. Pirogov, M. Artemev, Evaluating Deepfake Detectors in the Wild, *arXiv [cs.CV]* (2025); http://arxiv.org/abs/2507.21905.

**142**  W. Warby, Green Chameleon on a Branch (2024); https://unsplash.com/photos/IJAYYVG2V4Y.

**143\***  S. Gowal, R. Bunel, F. Stimberg, D. Stutz, G. Ortiz-Jimenez, C. Kouridi, M. Vecerik, J. Hayes, S.-A. Rebuffi, P. Bernard, C. Gamble, M. Z. Horváth, F. Kaczmarczyck, A. Kaskasoli, A. Petrov, I. Shumailov, M. Thotakuri, … P. Kohli, SynthID-Image: Image Watermarking at Internet Scale, *arXiv [cs.CR]* (2025); http://arxiv.org/abs/2510.09263.

**144**  J. Cao, Q. Li, Z. Zhang, J. Ni, Secure and Robust Watermarking for AI-Generated Images: A Comprehensive Survey, *arXiv [cs.CR]* (2025); http://arxiv.org/abs/2510.02384.

**145**  Y. Chen, Z. Ma, H. Fang, W. Zhang, N. Yu, TAG-WM: Tamper-Aware Generative Image Watermarking via Diffusion Inversion Sensitivity, *arXiv [cs.MM]* (2025); http://arxiv.org/abs/2506.23484.

**146**  T. South, Ed., "Identity Management for Agentic AI" (OpenID, 2025); https://openid.net/wp-content/uploads/2025/10/Identity-Management-for-Agentic-AI.pdf.

**147**  A. Chan, N. Kolt, P. Wills, U. Anwar, C. S. de Witt, N. Rajkumar, L. Hammond, D. Krueger, L. Heim, M. Anderljung, IDs for AI Systems, *arXiv [cs.AI]* (2024); http://arxiv.org/abs/2406.12137.

**148**  T. South, S. Marro, T. Hardjono, R. Mahari, C. D. Whitney, D. Greenwood, A. Chan, A. Pentland, Authenticated Delegation and Authorized AI Agents, *arXiv [cs.CY]* (2025); http://arxiv.org/abs/2501.09674.

**149**  European Commission, The General-Purpose AI Code of Practice. (2025); https://digital-strategy.ec.europa.eu/en/policies/contents-code-gpai.

**150**  S. Wiener, S. Rubio, *SB-53 Artificial Intelligence Models: Large Developers* (2025); https://leginfo.legislature.ca.gov/faces/billTextClient.xhtml?bill_id=202520260SB53.

**151**  G7, OECD, G7 Reporting Framework – Hiroshima AI Process (HAIP) International Code of Conduct for Organizations Developing Advanced AI Systems. (2025); https://transparency.oecd.ai/.

**151**  OECD, Launch of the Hiroshima AI Process (HAIP) Reporting Framework, *OECD* (2025); https://www.oecd.org/en/events/2025/02/launch-of-the-hiroshima-ai-process-reporting-framework.html.

**153**  K. Persec, J. Healy, S. F. Esposito, "Shaping Trustworthy AI: Early Insights from the Hiroshima AI Process Reporting Framework" (OECDAI, 2025); https://oecd.ai/en/work/haip-reporting-insights.

**154**  OECD, Submitted Reports – HAIP Reporting Framework (2025); https://transparency.oecd.ai/reports.

**155**  ASEAN, "Expanded ASEAN Guide on AI Governance and Ethics - Generative AI" (ASEAN, 2025); https://asean.org/book/

expanded-asean-guide-on-ai-governance-and-ethics-generative-ai/.

**156**  K. Choi, Analyzing South Korea's Framework Act on the Development of AI, *IAPP* (2025); https://iapp.org/news/a/analyzing-south-korea-s-framework-act-on-the-development-of-ai.

**157**  과학기술정보통신부, 인공지능 발전과 신뢰 기반 조성 등에 관한 기본법 (2025); https://www.law.go.kr/%EB%B2%95%EB%A0%B9/%EC%9D%B8%EA%B3%B5%EC%A7%80%EB%8A%A5%20%EB%B0%9C%EC%A0%84%EA%B3%BC%20%EC%8B%A0%EB%A2%B0%20%EA%B8%B0%EB%B0%98%20%EC%A1%B0-%EC%84%B1%20%EB%93%B1%EC%97%90%20%EA%B4%80%ED%95%9C%20%EA%B8%B0%EB%B3%B8%EB%B2%95/%2820676,20250121%29.

**158**  METR, Frontier AI Safety Policies (2025); https://metr.org/.

**159**  Frontier Model Forum, "Risk Taxonomy and Thresholds for Frontier AI Frameworks" (2025); https://www.frontiermodelforum.org/technical-reports/risk-taxonomy-and-thresholds/.

**160**  M. D. Buhl, B. Bucknall, T. Masterson, Emerging Practices in Frontier AI Safety Frameworks, *arXiv [cs.CY]* (2025); http://arxiv.org/abs/2503.04746.

**161**  METR, Forecasting the Impacts of AI R&D Acceleration: Results of a Pilot Study (2025); https://metr.org/blog/2025-08-20-forecasting-impacts-of-ai-acceleration/.

**162**  J. Wang, K. Huang, K. Klyman, R. Bommasani, Do AI Companies Make Good on Voluntary Commitments to the White House?, *arXiv [cs.CY]* (2025); http://arxiv.org/abs/2508.08345.

**163**  Future of Life Institute, "AI Safety Index: Summer 2025" (Future of Life Institute, 2025); https://futureoflife.org/wp-content/uploads/2025/07/FLI-AI-Safety-Index-Report-Summer-2025.pdf.

**164**  S. Campos, H. Papadatos, F. Roger, C. Touzet, O. Quarks, M. Murray, A Frontier AI Risk Management Framework: Bridging the Gap between Current AI Practices and Established Risk Management, *arXiv [cs.AI]* (2025); http://arxiv.org/abs/2502.06656.

**165**  I. Habli, R. Hawkins, C. Paterson, P. Ryan, Y. Jia, M. Sujan, J. McDermid, The BIG Argument for AI Safety Cases, *arXiv [cs.CY]* (2025); http://arxiv.org/abs/2503.11705.

**166**  J. Clymer, J. Weinbaum, R. Kirk, K. Mai, S. Zhang, X. Davies, An Example Safety Case for Safeguards against Misuse, *arXiv [cs.LG]* (2025); http://arxiv.org/abs/2505.18003.

**167**  J. Clymer, N. Gabrieli, D. Krueger, T. Larsen, Safety Cases: How to Justify the Safety of Advanced AI Systems, *arXiv [cs.CY]* (2024); http://arxiv.org/abs/2403.10462.

**168**  A. Goemans, M. D. Buhl, J. Schuett, T. Korbak, J. Wang, B. Hilton, G. Irving, Safety Case Template for Frontier AI: A Cyber Inability Argument, *arXiv [cs.CY]* (2024); http://arxiv.org/abs/2411.08088.

**169**  M. D. Buhl, G. Sett, L. Koessler, J. Schuett, M. Anderljung, Safety Cases for Frontier AI, *arXiv [cs.CY]* (2024); http://arxiv.org/abs/2410.21572.

**170**  M. D. Buhl, J. Pfau, B. Hilton, G. Irving, An Alignment Safety Case Sketch Based on Debate, *arXiv [cs.AI]* (2025); http://arxiv.org/abs/2505.03989.

**171**  B. Hilton, M. D. Buhl, T. Korbak, G. Irving, "Safety Cases: A Scalable Approach to Frontier AI Safety" (AI Security Institute, 2025); https://doi.org/10.48550/arXiv.2503.04744.

**172***  Anthropic, Anthropic's Responsible Scaling Policy, Version 1.0. (2023); https://www-cdn.anthropic.com/1adf000c8f675958c2ee23805d91aaade1cd4613/responsible-scaling-policy.pdf.

**173***  Google DeepMind, Frontier Safety Framework Version 3.0. (2025); https://storage.googleapis.com/deepmind-media/DeepMind.com/Blog/strengthening-our-frontier-safety-framework/frontier-safety-framework_3.pdf.

**174**  K. Perset, S. Fialho Esposito, "How Are AI Developers Managing Risks?" (OECD, 2025); https://doi.org/10.1787/658c2ad6-en.

**175**  L. Staufer, M. Yang, A. Reuel, S. Casper, Audit Cards: Contextualizing AI Evaluations, *arXiv [cs.CY]* (2025); http://arxiv.org/abs/2504.13839.

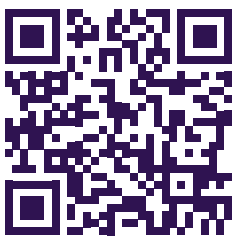

designbysoapbox.com